\documentclass{article}

\usepackage[final]{compsust_2023}
\usepackage{graphicx}
\usepackage{amsmath}
\usepackage{siunitx}
\usepackage[utf8]{inputenc} % allow utf-8 input
\usepackage[T1]{fontenc}    % use 8-bit T1 fonts
\usepackage{hyperref}       % hyperlinks
\usepackage{url}            % simple URL typesetting
\usepackage{booktabs}       % professional-quality tables
\usepackage{amsfonts}       % blackboard math symbols
\usepackage{nicefrac}       % compact symbols for 1/2, etc.
\usepackage{microtype}      % microtypography
\usepackage{xcolor}         % colors
\usepackage{wrapfig}

\title{Multi-fidelity Bayesian Optimisation of Syngas Fermentation Simulators}

\author{
  Mahdi Eskandari \\
  Division of Computing, Engineering and Mathematical Sciences\\
  University of Kent\\
  Canterbury, United Kingdom \\
  \texttt{me377@kent.ac.uk} \\
  \And
  Lars Puiman \\
  Faculty of Applied Sciences, Department of Biotechnology \\
  Delft University of Technology \\
  Delft, The Netherlands \\
  \texttt{L.Puiman@tudelft.nl} \\
  \AND
  Jakob Zeitler \\
  Matterhorn Studio \\
  Oxford, United Kingdom \\
  \texttt{jakob@matterhorn.studio} \\
}

\begin{document}

\maketitle

\begin{abstract}
A Bayesian optimization approach for maximizing the gas conversion rate in an industrial-scale bioreactor for syngas fermentation is presented. We have access to a high-fidelity, computational fluid dynamic (CFD) reactor model and a low-fidelity ideal-mixing-based reactor model. The goal is to maximize the gas conversion rate, with respect to the input variables (e.g., pressure, biomass concentration, gas flow rate). Due to the high cost of the CFD reactor model, a multi-fidelity Bayesian optimization algorithm is adopted to solve the optimization problem using both high and low fidelities. We first describe the problem in the context of syngas fermentation followed by our approach to solving simulator optimization using multiple fidelities. We discuss concerns regarding significant differences in fidelity cost and their impact on fidelity sampling and conclude with a discussion on the integration of real-world fermentation data.
\end{abstract}

% 4 PAGE MAXIMUM, one extra page for camera ready

\section{Introduction}
Syngas fermentation is a recently emerged technology, with a promise as a sustainable and circular process for the production of the fuels and materials from a wide range of feedstocks, such as biomass, municipal solid waste streams, and atmospheric $\mathrm{CO_2}$. LanzaTech, a prominent company in the field, has achieved successful commercialization of the fermentation process, wherein micro-organisms (\textit{Clostridium autoethanogenum}) convert synthesis gas (comprising $\mathrm{CO}$, $\mathrm{H_2}$, and $\mathrm{CO_2}$) from industrial off-gases into ethanol \citep{Fackler2021,Kopke2020}. While gas-to-liquid mass transfer limitations have historically posed challenges for scaling-up syngas fermentation, it is worth noting that such limitations can be significantly mitigated by producing bubble coalescence-suppressing products, such as ethanol  \citep{Puiman2022_CFD}.
Industrial-scale (syngas) fermentation process modeling has been established as a pivotal tool for reactor scale-up \cite{Delvigne2017}. The models used for such applications range from relatively simple, and fast, 0D and 1D models with a black-box description of the micro-organism to more complex 3D computational fluid dynamic (CFD) models (see e.g. \cite{Benalcazar2020,Siebler2019,puiman2023downscaling}). A detailed description of the microbial behavior in the reactor could be obtained by coupling reactor models with metabolic models (e.g. \cite{Chen2018,Haringa2018}). Although CFD models provide a high-resolution description of the bioreactor, solving CFD models takes considerably more time (order of days to weeks) than solving 0D and 1D models (seconds to minutes). This is a major limitation for assessing the impact and the optimization of operating conditions, such as biomass concentration, pressure or gas composition, as such an iterative solving procedure in a CFD model could take months. 
Here, we present an approach for optimization of operating conditions using a Bayesian optimization (BO) framework \citep{frazier2018tutorial}, and a 0D model, which was tested before on CFD modeling results \citep{puiman2023downscaling}. Specifically, we are using multi-fidelity BO, which allows us to flexible learn from both low and high-fidelity simulations \citep{poloczek2017multi}.

\section{Problem}
\label{headings}
\cite{puiman2023downscaling} present two bioreactor models for industrial-scale syngas fermentation: 

\begin{itemize}
    \item A \textit{high-fidelity}, CFD model, that takes a few ($\sim 5$) days per run.
    \item A \textit{low-fidelity}, ideal-mixing model that takes less than $2$ minutes per run.
\end{itemize}
 
 We are interested in finding the optimal operating parameters for industrial-scale syngas fermentation in these simulators, by efficiently exploring these two fidelities. 
The CFD model can be used to estimate the spatial variation in dissolved gas concentration and microbial uptake rates. It does so by solving the multiphase flow behavior in an Eulerian fashion, while the turbulence is solved for using a RANS (Reynolds-Averaged Navier-Stokes) model. Transport of CO, H$_2$, CO$_2$ is solved in both gas and liquid phases, while mass transfer and reaction are solved using the same methods as in the ideal-mixing model, but with high spatial distribution \citep{Puiman2022_CFD,puiman2023downscaling}. In the CFD model, all these equations are solved for $370,000$ mesh elements during a period of $1000$ s. One simulation usually lasts a week with $48$ AMD EPYC cores.   
The ideal-mixing model can be used for fast calculations of the industrial-scale syngas fermentation bioreactor. It is a good initial approximation (as observed in Figure 3 from \cite{puiman2023downscaling}) for the CFD model. Gas-to-liquid mass transfer and microbial syngas conversion are described by:
\begin{equation}
    (k_La)_i (H_ipy_i - c_{L,i}) = q_ic_X,
\label{Eq:Eq_idealMix}
\end{equation}
where 
\begin{align}
    (k_La)_i &= k_{L,i} \frac{6\epsilon_{G}}{d_B}, \hspace{2mm} \mathrm{for} \hspace{1mm} i = \mathrm{CO}, \mathrm{H_2}, \mathrm{CO_2}, \\
    \epsilon_{G} &= \frac{u_{G,s}}{0.25} ,\\
    u_{G,s} &= \frac{n_{G,s}RT}{Ap},
\end{align}
with the constants temperature $T = 273.15 K$, gas constant $R = 8.314 \frac{J}{\mathrm{
mol}K}$, reactor area $A = 19.63 m^2$, $k_{L, \mathrm{CO}} = 3.98 \times 10^{-4} m/s$, $k_{L, \mathrm{H_2}} = 5.93 \times 10^{-4} m/s$, $k_{L, \mathrm{CO_2}} = 3.87 \times 10^{-4} m/s$ \citep{puiman2023downscaling}. The typical parameters used for the ideal-mixing model are listed in Table~\ref{table-params}.

The following parameters were considered as a variable during operation:
\begin{enumerate}
    \item Biomass concentration $c_X \in [0, 50] g/L $
    \item Pressure $p\in [101325 - 506625] \mathrm{Pa}$
    \item Mole fraction $y_i$ where $i \in \{ 
 \mathrm
 CO_2, \mathrm{
 H_2}, \mathrm{
 CO}  \}$. Note that the constraint $\sum_i y_i = 1$ applies for the mole fractions.
 \item  Bubble diameter $d_B  \in [0, 10^{-3}] m$.
 \item Flow in gas inlet $n_{G,s} \in [0, 200 ] \mathrm{mol/s}$.
\end{enumerate}
With the given parameters, the output of the model is the $\mathrm{CO}$ and $\mathrm{H_2}$ reaction rate $r_i$, $i \in \{ \mathrm{CO},   \mathrm{ H_2}\}$ with $r_i = q_i c_X$. So the target is to calculate $q_i$, considering $\mathrm{CO}$ and $\mathrm{H_2}$ uptake kinetics \citep{puiman2023downscaling} and Equation \ref{Eq:Eq_idealMix}.
 \begin{align} \label{q_co}
     q_{\mathrm{CO}} &= q_{\mathrm{CO}}^{\max} \left(  \frac{c_{\mathrm{L, CO}}}{K_{\mathrm{S, CO}} + c_{\mathrm{L, CO}} + \frac{c_{\mathrm{L, CO}}^2}{K_I}  } \right) \\ \label{q_h2}
     q_{\mathrm{H_2}} &= q_{\mathrm{H_2}}^{\max} \left( \frac{ c_{\mathrm{L, H_2}}  }{ K_{\mathrm{S, H_2}} + c_{\mathrm{L, H_2}} }    \right) \left(  
 \frac{ 1 }{ 1 + \frac{c_{\mathrm{L, CO}}}{K_{\mathrm{I, CO}}} }   \right).
 \end{align}
The parameters for solving (\ref{q_co}) and (\ref{q_h2}) are listed in Table~1 in \citep{puiman2023downscaling}. The CFD (high fidelity) and ideal-mixing (low fidelity) model both take the same values as input and give $q_i$ $i \in \{ 
  \mathrm{
 H_2}, \mathrm{
 CO}  \}$ as output. The comparison of the results of both simulators, for varying biomass concentrations $c_X$ is provided in Figure~3 in \citep{puiman2023downscaling}. The difference between them as mentioned is that the high fidelity model takes more than two days to complete a run whereas the scale-down simulator takes less than a minute to provide  results since its only complexity comes from solving a system of equations in (\ref{q_co}) and (\ref{q_h2}). The ideal-mixing model, however, is a good estimation for the average $q_{CO}$ and $q_{H2}$, while the CFD model also provides information on its spatial distribution inside the bioreactor. 
 The goal is to maximize the syngas conversion rate $r_i$ with respect to the input values. Since the industrial-scale simulator has a high complexity and cost for each experiment, directly maximizing its objective function is not feasible. 
 
\begin{figure}
    \centering
    \includegraphics[scale=0.11]{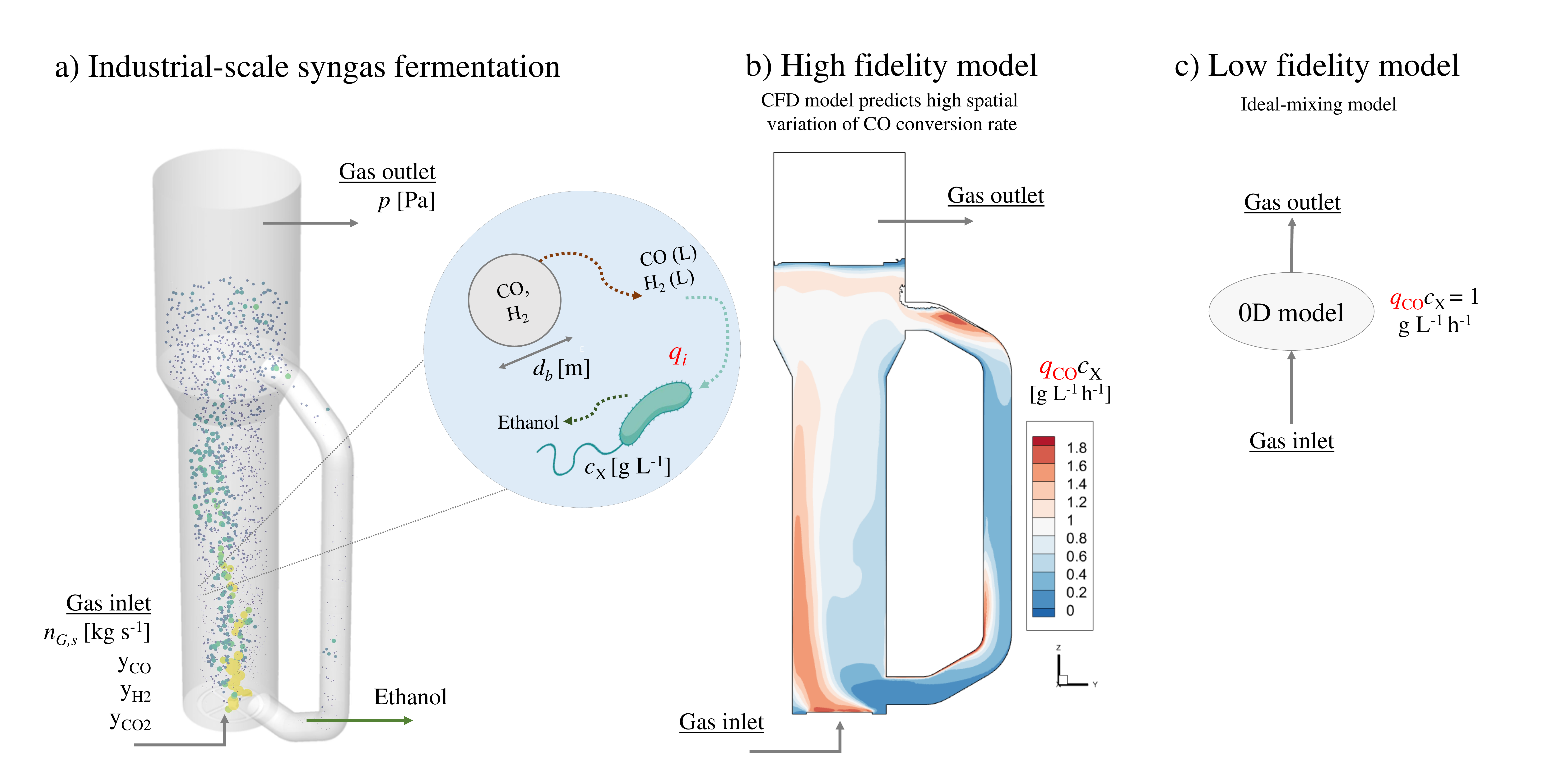}
    \caption{Schematic descriptions of the used models. a) Conceptual representation of the industrial-scale reactor, and the operating parameters, b) result of a run of the high fidelity CFD model, c) result of the low-fidelity ideal-mixing model. The variable in red $q_i$ is the one solved for in both modeling routines. }
    \label{fig:enter-label}
\end{figure}

\section{Solution}
\label{others}
In order to solve the above optimization problem, we apply multi-fidelity Bayesian optimization (MFBO, \citep{poloczek2017multi}). Classic Bayesian optimization (BO, \citep{frazier2018tutorial}) is a machine learning technique for efficiently optimizing complex functions, particularly when evaluations are computationally expensive and there is no access to a gradient. It combines Bayesian statistical modeling with information-gain functions to iteratively refine its modeling of the objective function to guide the search for minima or maxima. Usually, BO is started with a small set of evaluations to build a \textit{probabilistic model }(often a Gaussian Process, \citep{RasmussenW06}) that captures the function's behavior and uncertainty around model predictions. It then selects the next point to evaluate based on \textit{acquisition functions}, which, depending on operator preference, explore uncertain regions or exploit promising areas. This optimisation loop continues iteratively, gradually converging towards the optimal solution(s) while trying to keep small the number of function evaluations required. BO is widely used in machine learning hyperparameter tuning, experimental design, and other domains where optimizing costly functions is a challenge.
Since there are two fidelities available, i.e., high fidelity (the CFD model), and low fidelity (the ideal-mixing model), discrete MFBO \citep{poloczek2017multi} is utilized to solve the problem. The goal of \citep{poloczek2017multi} is to maximize an expensive-to-evaluate function where a cheaper approximation of the black box is also available. The high fidelity model is modeled as $\mathcal{IS}_0$ whereas the low fidelity one is modeled as $\mathcal{IS}_1$. Each of these information sources is also associated with a cost function $c_j$ with $j = \{0, 1 \}$. It is assumed that the cost function is known and continuous. For instance, in the syngas fermentation model, the cost function can be modeled as the time required for each run. 
The optimization algorithm proceeds in rounds such that using the previously sampled data a posterior Gaussian process is fit to the initial data. 
Then,
in each run an information source is picked that maximizes the cost-sensitive
Knowledge Gradient in the equation in  (\ref{fig:mkg_acqu}). Then with the new observation, the posterior is updated. The details of the misoKG algorithm are listed in \citep{poloczek2017multi} and are omitted for brevity. 
\begin{equation}
\label{fig:mkg_acqu}
    \mathrm{MKG}^n(l,x) = \mathbb{E}_n\left[ \frac{\max_{x^\prime \in \mathcal{D}} \mu^{(n+1)}(0, x^\prime) - \max_{x^\prime \in \mathcal{D}} \mu^{(n)}(0, x^\prime)}{c_l(x)} \big| l^{(n+1)} = l, x^{(n+1)} = x \right].
\end{equation}
\section{Discussion}

Before we commit to an optimization campaign of unknown length, we wanted to understand the MFBO behavior better to adjust our expectations on how many high-fidelity simulator runs (each 2-5 days) we should expect. Eq. (\ref{fig:mkg_acqu}) shows the equation of the acquisition function used in our MFBO model. Most simply, it is a nested optimization problem, where we maximize the gain in the objective function for each information source $\mathcal{IS}_l$ and then choose the source with the highest gain overall. The MKG for each information source is weighted by the cost of acquiring a sample from it.

We have significant concerns regarding a situation in which low-fidelity queries would dominate, resulting in fewer requests for high-fidelity runs. This could lead to a cubic increase in the demand for modeling resources, posing a significant computational obstacle. A brief examination of Eq.~\ref{fig:mkg_acqu} from \citep{poloczek2017multi} reveals that when the cost $c_l(x)$ decreases, the MKG value increases. This suggests that the model is more inclined to recommend that particular point as the next sample query. In fact, from an intuitive standpoint, if a low-fidelity information source is significantly less expensive than its high-fidelity counterpart while offering comparable performance, it is logical to predominantly depend on the low-fidelity source for experimentation.

Informed by the equation in Eq.~\ref{fig:mkg_acqu}, we did a brief empirical study to understand the influence of fidelity cost on information source selection. Presumably, the lower cost of the low fidelity will cause the acquisition function to suggest more queries of the low fidelity. We ran the standard BoTorch discrete MFBO tutorial \citep{botorch_tutorial}  on optimizing a multi-fidelity version of the Hartmann6D test function with costs of the lower fidelity set at $c_l(x) \in \{0.001, 0.05, 0.1\}$, 8 times each for different seeds, for 50 acquisition steps. We are interested to see how different low-fidelity costs impact the budget spend, both low and high fidelity added together, in the first 50 acquisition steps.
\begin{figure}
    \centering
    \includegraphics[scale = .4]{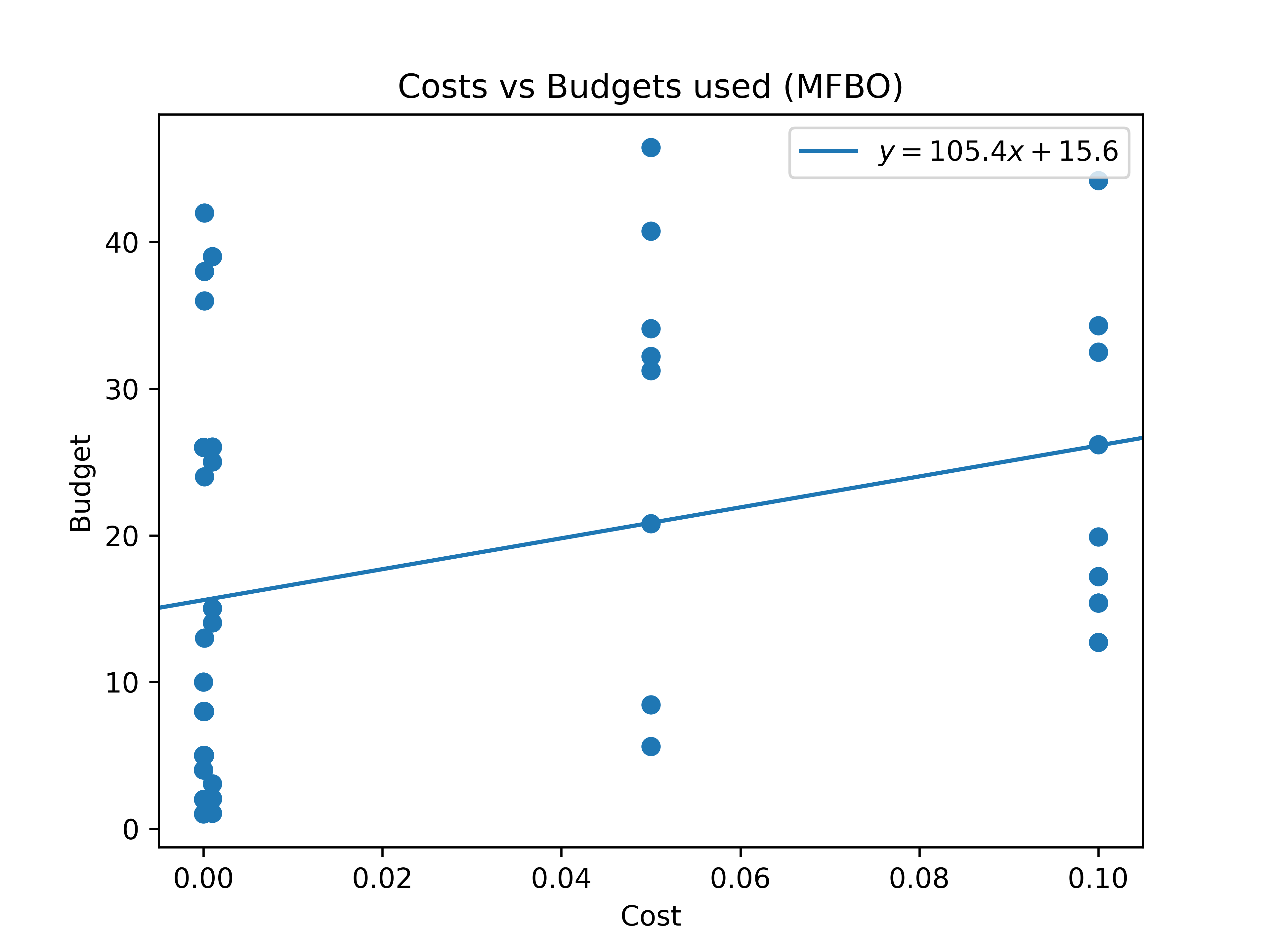}
  \caption{Linear regression of budget spend onto the cost of the single fidelity. The linear relationship is shown top right. The slope is statistically significant at a $p$-value of $0.038$.}
    \label{fig:costsvsbudgets}
\end{figure}
Based on those samples, we then ran a Pearson correlation test and linear regression of the budget used for each trial and the cost $c_l(x)$. The hypothesis is that as the cost of the low fidelity increases, so does that budget-use over $50$ iterations. We identified a positive correlation with a $p$-value of $0.038$ which generally is accepted as statistically significant and as such is in agreement with both theory and intuition laid out above, see Figure \ref{fig:costsvsbudgets}.
Equipped with the intuition on how MFBO will query low and high-fidelity via the above empirical study, we are now confident as a next step to engage in an MFBO campaign to optimize the high-fidelity simulator with the help of low-fidelity samples. A possible extension would be a modification of the acquisition function would tie the execution time of the two fidelities into the cost and down-weight the querying of low-fidelities.

Ultimately, we hope to incorporate real-world fermentors as supplementary high-fidelity information sources alongside the existing high-fidelity simulator. Fermentation engineers can utilize the publicly available MFBO code \footnote{See tutorial and single-click pipeline access here: \url{https://matterhorn.studio/pages/seminars/syngas-fermentation-optimisation/}} to optimize their fermentation conditions. They can pool both low and high-fidelity simulators such as in \citep{puiman2023downscaling}, as well as their large-scale fermentation data, to maximize the reaction rate of their syngas bioreactors.

\section{Conclusion}
We presented a multi-fidelity Bayesian optimization approach for the optimization of syngas conversion in industrial-scale bioreactors. With real-world experimentation often prohibitively expensive, simulation approaches are an indispensable choice for bioreactor engineers tasked with scale-up. Pooling the data between simulators of different fidelity and costs can make a significant difference to fermentation engineering. We anticipate that our computational approach presented here can guide the optimization search in both academia and industry. We hope that it will help identify the optimal operating conditions for bioreactors as part of the critical contribution that syngas fermentation can make toward a sustainable planet.

\section{Acknowledgments}
Thank you to the workshop reviewers for their detailed feedback that allowed us to improve and clarify our work. We are also grateful for the support for this work from the UKRI Innovate UK Transformative Technologies Grant 2023 Series.

\section*{Appendix: List of symbols}
\begin{table}[h]
\caption{List of the parameters used for simulating the ideal-mixing model.}
\label{table-params}
\centering
\small
\begin{tabular}{|l|p{6cm}|p{4cm}|}
\hline
Symbol      & Description                               & Unit \\
\hline
$A_p$       & Reactor area                              & $\si{\square\meter}$ \\
$c$         & Concentration in liquid phase             & $\si{\gram\per\liter}$ or $\si{\mol\per\cubic\meter}$ \\
$d_B$       & Bubble diameter                           & $\si{\meter}$ \\
$\epsilon_G$ & Gas hold-up                             & - \\
$H$         & Henry constant                            & $\si{\mol\per\cubic\meter\per\pascal}$ \\
$K_i$       & Inhibition constant                       & - \\
$k_L$       & Mass transfer coefficient                 & $\si{\meter\per\second}$ \\
$k_La$      & Volumetric mass transfer coefficient      & $\si{\per\second}$ \\
$K_s$       & Half-saturation constant                  & - \\
$n_{G,s}$   & Molar gas flow                            & - \\
$p$         & Pressure                                  & $\si{\pascal}$ \\
$q$         & Biomass-specific rate                     & $\text{mol}_i \, \text{mol}_X^{-1}\,\si{s^{-1}}$ \\
$R$         & Gas constant                              & $\si{\joule\mol\kelvin}$ \\
$T$         & Temperature                               & $\si{\kelvin}$ \\
$u_{G,s}$   & Superficial gas velocity                   & $\si{\meter\per\second}$ \\
$y$         & Gas fraction                              & $\text{mol}_i \, \text{mol}_G^{-1}$ \\
$i$         & Compound                                  & - \\
$L$         & Liquid phase                              & - \\
$X$         & Biomass                                   & - \\
\hline
\end{tabular}
\end{table}

\bibliographystyle{unsrtnat}
\footnotesize
\bibliography{refs}

% {
% \small

% [1] Alexander, J.A.\ \& Mozer, M.C.\ (1995) Template-based algorithms for
% connectionist rule extraction. In G.\ Tesauro, D.S.\ Touretzky and T.K.\ Leen
% (eds.), {\it Advances in Neural Information Processing Systems 7},
% pp.\ 609--616. Cambridge, MA: MIT Press.

% [2] Bower, J.M.\ \& Beeman, D.\ (1995) {\it The Book of GENESIS: Exploring
%   Realistic Neural Models with the GEneral NEural SImulation System.}  New York:
% TELOS/Springer--Verlag.

% [3] Hasselmo, M.E., Schnell, E.\ \& Barkai, E.\ (1995) Dynamics of learning and
% recall at excitatory recurrent synapses and cholinergic modulation in rat
% hippocampal region CA3. {\it Journal of Neuroscience} {\bf 15}(7):5249-5262.
% }

%%%%%%%%%%%%%%%%%%%%%%%%%%%%%%%%%%%%%%%%%%%%%%%%%%%%%%%%%%%%

\end{document}